# HiPIMS magnetized plasma afterglow diagnostic

M. Ganciu*, B. Butoi, A. Groza,  B. Mihalcea

National Institute for Laser, Plasma and Radiation Physics, PO Box MG-36, 077125, Magurele, Bucharest, Romania,

*mihai.ganciu@inflpr.ro

**Abstract**

Deposition of thin films that are resistant to aggressive environment conditions, using the High-power impulse magnetron sputtering (HIPIMS) technique, represents a technological challenge. To establish the optimal operating conditions it is desirable to know the lifetime of the magnetized plasma at the end of the pulsed cathode current. The table-top system we developed allows one to measure the lifetime for all operating HiPIMS types, both for short and long pulses, based on the rapid increase of the current when applying a voltage pulse with very short rise time, up to a value which depends of the initial magnetized plasma characteristics.

1. Introduction

Currently, use of plasma deposition techniques to achieve thin films that resist under aggressive environment conditions is a technological challenge. To deposit dense and ultra-dense films, the High-power impulse magnetron sputtering (HIPIMS) technique is employed, which ensures highly ionized pulverised vapours and a high molecular dissociation rate [1]. Many studies are developed with respect to phenomena associated with magnetized-plasma high ionization, which leads to a high energy impulse deposition on the cathode surface (1÷3 KW/cm$^2$), with an average deposited energy 1÷10 W/cm$^2$ on the cathode [2]. An interesting application is high quality optical deposition [3], and recently it was demonstrated that by using a short-pulse HiPIMS technique one can obtain thin films with superior characteristics, that are able to resist under extreme conditions characteristic to the cosmic space [4]. DLC coatings by HiPIMS technique [5] could also represent a viable alternative for space applications [6]. The obvious advantages of using  ultra-short pulses are associated with obtaining high quality deposited films: denser films, with a good adhesion and lower thermal load of the substrate, a better coating of complex surfaces preventing the transition to self sputtering, stable and reproducible reactive processes, relaxation times of the deposited surfaces that are compatible with ion packages deposited in short pulses, etc. [7, 8, 9] .

2. Experimental

The response of the HiPIMS equipment to intense cathode current pulses is essential for identifying/realizing adequate electric HIPIMS systems. The pre-ionisation turns out to be useful in reducing the building time of the magnetized plasma that controls the cathode current for an optimal cathode voltage. The initial, low-density magnetized plasma, can be obtained by means of DC discharges, pulse succession and RF discharges [7]. Experimental devices developed (Figure 1) include classical magnetrons with rectangular or circular cathodes made from different metals: titanium, copper, ruthenium etc. The power supply and discharge plasma are considered as a coupled system [10]. An adapted pulse generator is used to apply high power pulses (1 ÷ 50 µs, 10 ÷ 200 A, 500 ÷ 1200 V, 0÷ 1 kHz) directly on the cathode, thus insuring a short fall time ( < 1 µs) of the magnetron current pulses [9 -18]. It also allows the superposition of an additional pre-ionization discharge.

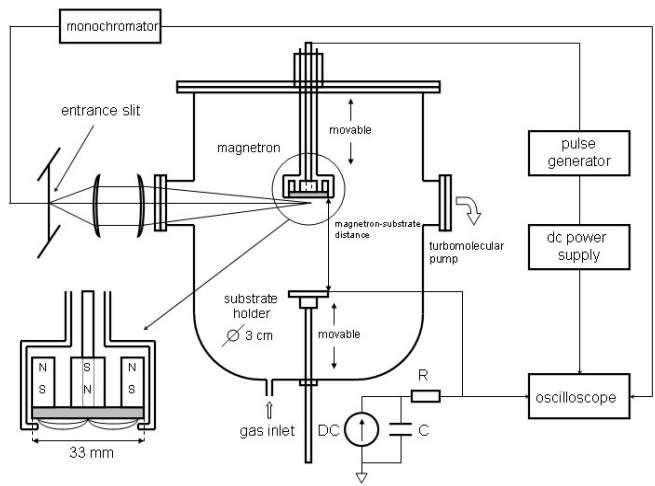

*Figure 1: Developed pulsed magnetron experimental devices: self-sputtering of copper [9]*

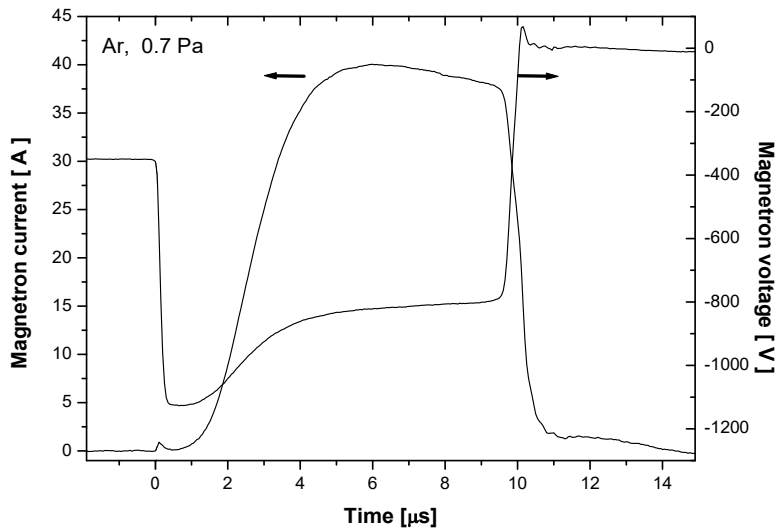

*Figure 2: Typical cathode voltage and current waveforms*

Figure 2 illustrates a typical applied voltage pulse at the cathode (target), and the magnetron current for argon working gas at 1.4 Pa and a DC preionization current of 10mA. We obtain a maximum current density of 1 to 20 A/cm$^2$, over a pressure range that lies between 0.4 ÷ 5 Pa, for different cathode geometries. For high cathode current densities, the cathode plasma kinetic pressure can exceed the magnetic pressure. Taking into account only the electron contribution $n_e k T_e$ and assuming an electronic temperature ~ 3 eV, a simple estimation for B = 200 gauss and J = 4A/cm$^2$ yields a kinetic pressure higher than the magnetic one. The electron density is estimated from the ion density current by using the Bohm model [22]. Thus, we can suppose that at high cathode current densities (which also implies high electronic densities) we can facilitate plasma expansion by magnetic plasma deconfinement [11, 12]. Moreover, as the ions in self-sputtering mode result from vapour ionization, we can assume that the ion temperature is supplied by the average energy of sputtered vapour (~ 3 eV). If the plasma kinetic pressure is much larger than the magnetic pressure, generally the plasma pushes the magnetic field around and carries it along with its natural motion. An example of a simple qualitative simulation of magnetic field perturbation induced by the cathode dense plasma in expansion due to its diamagnetic properties [11], is presented in Figure 3.

Two physical phenomena are mainly responsible for the deposition rate reduction in HiPIMS with respect to DC magnetron sputtering. First, the self-sputtering transition with low self-sputtering yield of titanium atoms can be invoked. Second, metal ions must be able to diffuse freely towards the substrate and the space charge of the ion flux must be compensated.

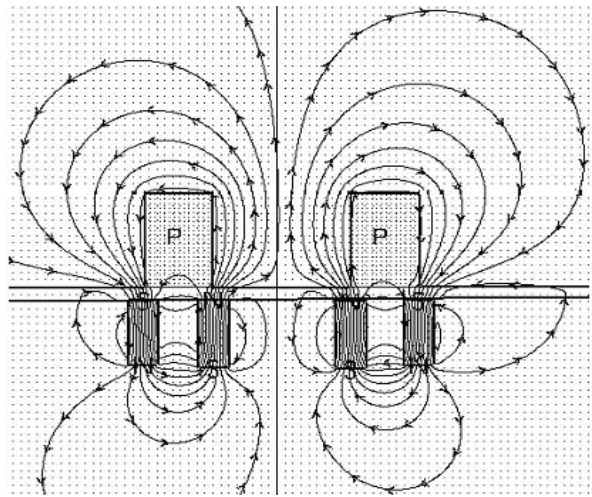

*Figure 3: Simulation of magnetic field perturbation by the presence of high density plasma (P). In this simulation for the diamagnetic effect of the dense plasma we assume plasma relative magnetic permeability of 0.1. The magnetic field in the absence of the plasma is shown in the bottom of the figure.*

Another approach aimed at increasing the deposition rate by improving the recuperation of the ionized vapour accumulated in magnetized plasma, lies in shortening the fall time of the magnetron current and applying an inverse pulse to the cathode [https://arxiv.org/abs/1906.09402]. Therefore, a larger percentage of the accumulated ions in magnetized plasma are directed to the substrate. Results obtained for a compact table-top pulsed magnetron developed for large pressure range operation [13], show an ion current increase just after the magnetron current is turned-off as illustrated in Figure 4, validating the importance of the afterglow magnetised plasma.

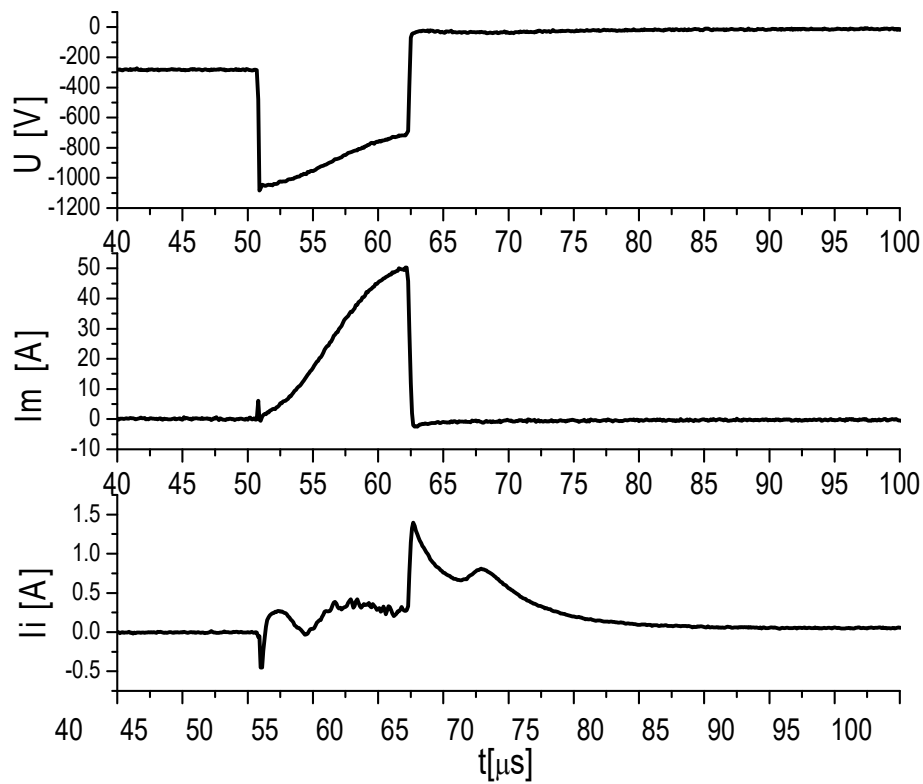

*Figure 4: Cathode voltage (U), magnetron current (Im) and substrate ionic current (Ii) for 300 mTorr Ar pressure, 3 cm diameter of the copper cathode, 1.5 cm cathode-substrate distance, and -40 V substrate bias voltage*

The compact configuration of the device allows a control of the plasma confinement with an external magnet near the substrate (Figure 5)

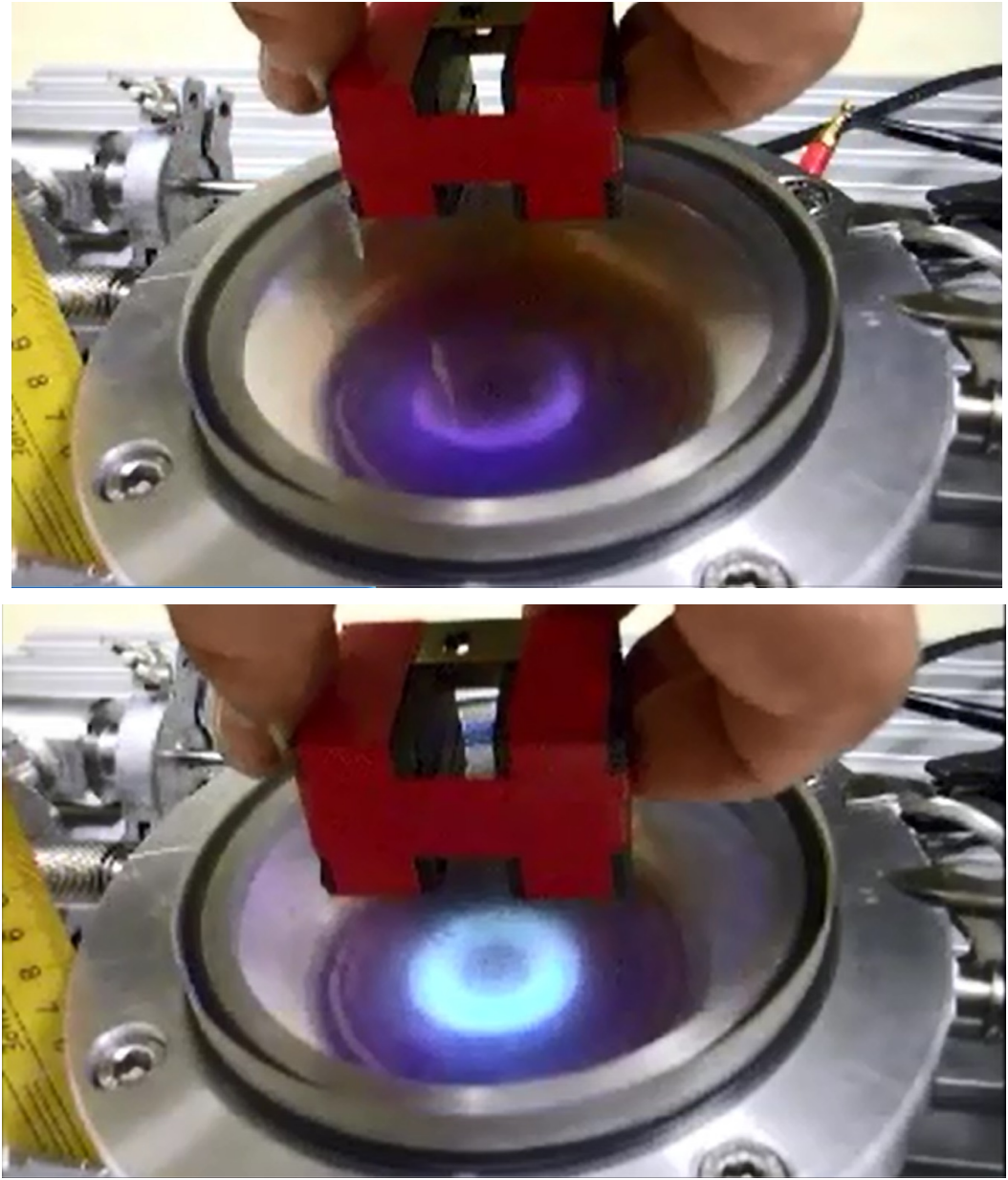

*Figure 5: Control of the plasma confinement with external magnets near the substrate*

To establish the optimal operating conditions, it is helpful to know the lifetime of the magnetized plasma at the end of the cathode pulsed current. The table-top system developed by us allows one to measure this time for all operating HiPIMS types, both for short and long pulses, based on the rapid increase of the current when applying a voltage pulse with very short rise time, up to a value which depends of the magnetized plasma characteristics

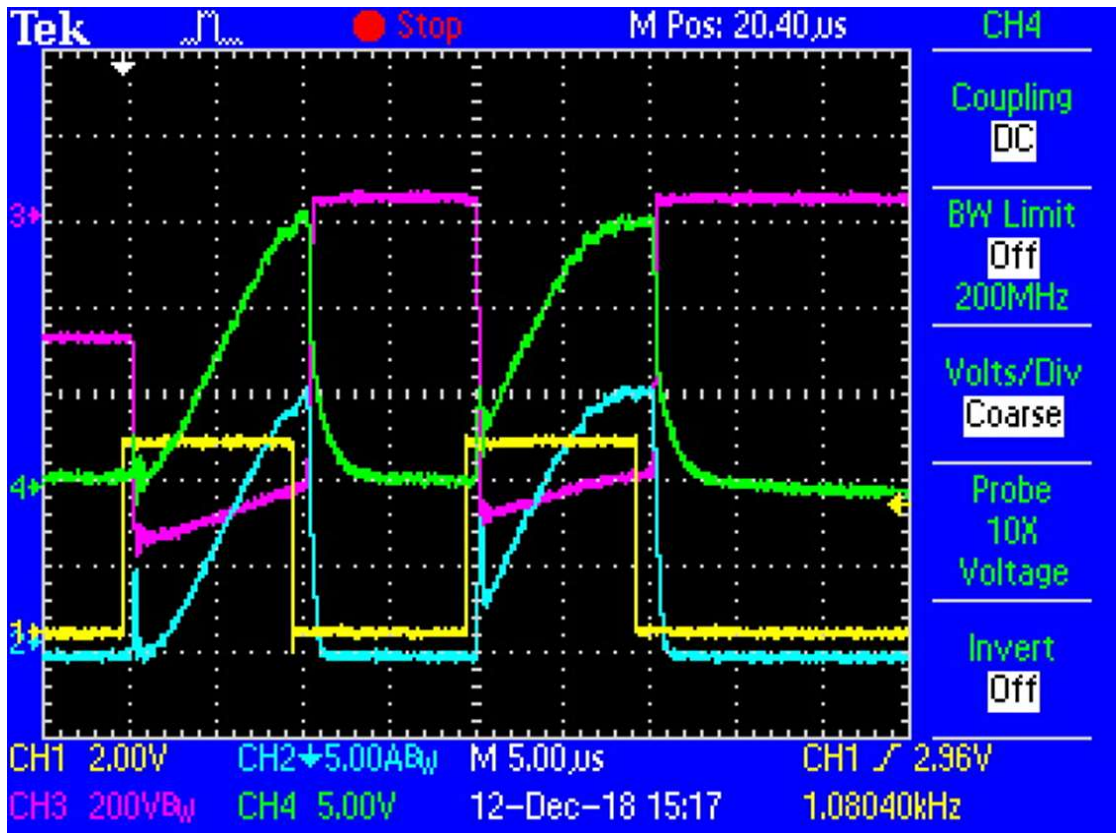

*Figure 6: Double pulse operation of a HiPIMS from UAIC-Iasi*

*Cathode voltage (pink), magnetron current (blue)*

The magnetron current rise time is limited by the electrical characteristics of the pulse generator only at the beginning, when magnetised plasma caused by the preceding pulse is still present. After this first rise of the current, any further rise of the current depends on the time characteristics of the magnetized plasma building. By modifying the delay time between the first pulse and second pulse, we can measure the exponential decay characteristic to the magnetized plasma, with a maximum value at the end of the main pulse:

https://www.youtube.com/watch?v=Mag1SF43f8Q

This decay time depends on the magnetron characteristics and allows us to define the operating parameters for the input system in HiPIMS regime for various applications, particularly for dedicated optical components and specific targets used in high power laser experiments [15,16].

3. **Conclusions**

A reverse pulse applied to the cathode can prevent deposition of accumulated ions in the magnetized plasma, and then most of them will be found on the substrate. It is possible to set up the optimal operating frequency in such a way that it does not overlap the pulses over the magnetized plasma generated by the preceding ones. In contrast, if it overlaps partially it ensures preionization without the need of an additional source. The dynamics of the afterglow plasma depends on the magnetron geometry, the discharge parameters, etc., and it can represent an object of fundamental process investigation, especially in the reactive mode. Complex compounds in the afterglow are not destroyed by energetic electrons that are firstly lost.

**Conflicts of Interest**: The authors declare no conflict of interest. The funders had no role in the study layout and design; in the collection, analysis, or interpretation of data; in writing the manuscript, or in the decision to publish the results.

**Acknowledgments**: We acknowledge fruitful discussions with Prof. Gheorghe Popa and Dr. Vasile Tiron from UAIC-Iasi and Dr. Catalin Vitelaru from INOE – Magurele,

**Funding**: This work has been funded by European Space Agency within the ESA contract No.4000121912/17/NL/CBi, by Romanian National Authority for Scientific Research and Innovation, contract No. 3N/2018 (Project PN 18 13 01 03) and by Romanian Space Agency, contract No. 53/19. 11. 2013 (Competence Center: Laser-Plasma Acceleration of Particles for Radiation Hardness Testing - LEOPARD)